\documentstyle[aps,12pt]{revtex}

\begin{document}

\title{Transverse Radial Expansion and Directed Flow} 

\author{Sergei A. Voloshin\footnote{On leave from Moscow 
Engineering Physics Institute, Moscow, 115409,  Russia}}

\address{Physikalisches Institut der Universit\"{a}t Heidelberg, 
Heidelberg, Germany}

\date{\today}
\maketitle

\begin{abstract}           
The effects of an interplay of radial expansion of the thermalized 
system created in a heavy ion collision and directed flow are discussed.
It is shown that the study of azimuthal anisotropy of particle distribution 
as a function of rapidity {\em and} transverse momentum could reveal 
important information on both radial and directed flow.
\end{abstract}


\narrowtext

Recently, the  study of collective flow in nuclear collisions 
at high energies has attracted an increased attention of both theoreticians 
and experimentalists. 
There are several reasons for that: 
i) the observation of anisotropic flow at 
the AGS~\cite{l877flow1,l877flow2} and, probably, at the SPS~\cite{lna49} 
energies, 
ii) better theoretical understanding of the relation  
between appearance and development of flow pattern during 
the fireball evolution and the processes such as thermalization, creation 
of quark-gluon plasma,  phase transitions, etc.~\cite{ljs,lbrav,lko2,lshur},
iii) the study of mean field effects~\cite{lko2,lsorge},
iv) the importance of flow for other measurements such as two-particle
interferometry~\cite{lheinz,lheinz2,lcsorgo,lvc,lmisk}, 
v) development of new techniques suitable for
flow study at high energies~\cite{lvzh,lolli}.
Although all forms of flow are interrelated and represent only different 
parts of one global picture, usually people discuss different forms 
of collective flow, such as longitudinal expansion, radial transverse 
expansion, directed flow, elliptic flow~\cite{lolli,lsorge}.

Flow introduces strong space-momentum correlation in the particle
production. 
Particles with a given rapidity and transverse momentum are produced 
only by some part of the entire source. 
This part we call as an effective source. 
It moves with the rapidity close to that of 
the particles~\cite{lheinz,lcsorgo} in the midrapidity
region, or slightly less than that for particles in the fragmentation 
region~\cite{lmisk}.
The subject of the current study is transverse (radial and directed) flow, 
namely the questions how azimuthally symmetric radial flow interferes with
directed flow and affects the directed flow signal.

It is convenient to characterize directed flow by $v_1$,
the amplitude of the first harmonic in the Fourier decomposition 
of the particle azimuthal distribution~\cite{lvzh}.
Depending on which particular distribution is studied, the coefficients
in the Fourier decomposition depend on different variables.
If it is a 2-dimensional rapidity and azimuthal angle distribution,
then $v_1$ depends only on rapidity, and 
$v_1=\langle p_x \rangle /\langle p_t \rangle$, where $\langle p_t \rangle$ 
is the mean transverse momentum, and $\langle p_x \rangle$ is 
an another widely used quantity to describe flow, the mean projection of 
the transverse momentum onto the reaction plane.
If one studies the production of particles with a given rapidity
and transverse momentum, then
\begin{equation}
E \frac{d^3 N}{d^3 p} = 
\frac{d^3N}{p_t d p_t dy d\phi'}=
\frac{1}{2\pi} \frac {d^2N}{p_t d p_t dy}(1+2 v_1 \cos (\phi')
+2 v_2 \cos(2 \phi')+...),
\end{equation}
and $v_1$ depends on two variables $y$ and $p_t$.
The dependence $v_1(p_t)$ for particles with a given
rapidity is the main topic of the current study. 
We will show that the radial expansion of the source results in specific
shapes of this function, which can be studied experimentally, 
providing an information on the directed as well as radial components of
the effective source velocity.
 
First studies of event shape anisotropy as a function of transverse
momentum~\cite{lyzh,lko1} have shown very interesting results. 
Such a study using Fourier decomposition of the azimuthal distribution 
promises better quantitative description of the effect.
One of the advantages would be the possibility to correct 
the results for the reaction plane resolution~\cite{l877flow2,lvzh}.
The analysis of $v_1(p_t)$ is to a large extend independent from the
uncertainties in the $p_t$ dependence of spectrometer acceptance and 
efficiency, and in this sense such an analysis has preference in comparison
with the analysis of triple differential distributions.

Transverse directed flow is a result of a  movement of   
an effective source in the transverse plane
(below we assume that this movement is along the ``x'' axis). 
In the source rest frame the first moment of azimuthal distribution 
($v_1$) is zero.
The final anisotropy (here and below we discuss only the first harmonic of
the azimuthal distribution) appears only as a consequence of the source 
movement in the transverse direction. 
Below we derive a general expression for $v_1$, assuming
that the invariant distribution in the source rest frame  is known
and (for simplicity) azimuthally symmetric:
\begin{equation}
E \frac{d^3N}{d^3p} \equiv J({\bf p_t},y)=J(p_t,y).
\end{equation} 
Source movement in the transverse direction results in
$J \rightarrow J'=J({\bf p_t}',y)$, where ${\bf p_t}'$ is defined 
by the Lorentz shift with velocity $\beta_a$ along the ``x'' axis. 
Assuming that the shift is small (the directed flow velocity expected
to be $\beta_a \leq 0.1$), $p_x'=p_x-\beta_a E$, and
taking into account that $\partial p_t/ \partial p_x=\cos(\phi)$, one gets:
\begin{equation}
v_1=\frac{\int d\phi J({\bf p_t}',y) \cos(\phi)}{\int d\phi J({\bf p_t}',y)}=
     -\frac{\beta_a E}{2} \frac{d \ln J(p_t,y)}{dp_t}.
\label{ev1}
\end{equation}
A comparison of exact numerical calculations for the models discussed below
with the results of application
of this formula shows that for the values of $\beta_a \leq 0.1$, 
the formula~(\ref{ev1}) is accurate at the level of a few percent.

Below we apply the derived formula to a few particular cases,
when the thermal source undergoes an isotropic expansion.
We first consider an expansion in a non-relativistic case, then
go to a relativistic generalization, and, finally, discuss a model
of isotropically expanding thermal shell~\cite{lre}, 
widely used in the analysis of BEVALAC and SIS data~\cite{leos,lnh}.  
Using simple models permits us to perform all calculations analytically 
to keep clear the qualitative features of the effect.
Note that directed as well as isotropic expansion velocities are
defined in the frame moving longitudinally with the rapidity of the effective
source.

Directed flow of protons is the most pronounced and less affected by
another effects such as shadowing.
Protons are relatively heavy particles ( $m \gg T$, $m$ is the proton 
mass, $T$ is the temperature), and often can be treated 
as non-relativistic particles. 
This is certainly justified in the source rest frame, 
where the proton kinetic energy is of the order of temperature 
$E^*-m \sim T \ll m$.
To have possibility to treat protons non-relativistically in the transverse 
direction in the {\em analysis} frame we have to assume that $m_t-m \ll m$,
what restricts the non-relativistic consideration to the 
region of $p_t \leq 0.5~GeV$.

The transverse isotropic expansion of the source can be described
as a superposition of different sources moving radially with an expansion
velocity $\beta_0$. 
Then the (non-relativistic) transverse momentum distribution of protons
from a radially expanding thermal source can be written as:
\begin{equation}
\frac{1}{N}\frac{d^2N}{d{\bf p_t}}=\frac {1}{(2\pi)^2 \, (2m T)} 
\int d\psi 
\exp\left(-\frac{(p_x-p_0 \cos(\psi))^2+(p_y-p_0 \sin(\psi))^2}{2m T}
\right),
\end{equation}
where $p_0=m \beta_0$.
The integration over $\psi$ (the orientation of the expansion velocity)
results in the distribution:
\begin{equation}
\frac{1}{N} \frac{d^2N}{d{\bf p_t}}=\frac{1}{2\pi (2m T)} 
\exp\left(-\frac{p_t^2+p_0^2}{2m T}\right) I_0(\xi),
\end{equation}
where $I_0$ is the modified Bessel function, and $\xi=\beta_0 p_t / T$.

Using the formula~(\ref{ev1}) one gets an expression for $v_1$:
\begin{equation}
v_1(p_t)=\frac{p_t \beta_a}{2T}
      \left(1-\frac{m \beta_0}{p_t} \frac{I_1(\xi)}{I_0(\xi)}\right). 
\label{v1nr}
\end{equation}
In our analysis we work in the frame moving longitudinally
with the effective source rapidity.
In this frame the longitudinal particle momenta usually can be neglected, 
what was done in the derivation above. 
If one considers the particle production with rapidities far from the
rapidity of the effective source, when particle longitudinal momenta are 
large, one should make a substitution $m \rightarrow \sqrt{m^2+p_l^2}$.

It follows from (\ref{v1nr})
that in the case without radial expansion $v_1$ linearly depends
on the transverse momentum with the slope of $0.5 \beta_a/T$.
The radial expansion of the system decreases the directed flow signal
changing also the shape of $v_1(p_t)$ in the low $p_t$ region.
Remarkably, that at some parameter values, namely such that 
$m \beta_0^2 /2 > T$ (for $T=100~MeV$ it corresponds to $\beta_0 > 0.46$), 
$v_1(p_t)$ has a region (at small $p_t$), 
where it is negative (see Fig.~1).
Physically it corresponds to the case, when particle production
with such a value of $p_t$ is more probable from the part of the effective
source which moves to the opposite direction than the flow direction (in this
case directed flow and ``expansion'' flow compensate each other).

The relativistic generalization of the above formulae is strait-forward,
although results into somewhat less transparent expressions.
The invariant distribution in this case can be written in the form:
\begin{eqnarray}
J &\propto& \int d\psi E^* e^{-E^*/T} 
\nonumber\\
  &=&\int d\psi (E \cosh(y_t) -p_t \cos(\psi) \sinh(y_t))
\exp(-(E \cosh(y_t) -p_t \cos(\psi) \sinh(y_t))/T)
\nonumber\\
&=&T e^{-\chi}(\chi I_0(\xi) - \xi I_1(\xi)),
\label{ej}
\end{eqnarray}
where $\chi=E \cosh(y_t) /T$, $\xi=p_t \sinh(y_t) /T$; 
 $y_t=0.5 \ln ((1+\beta_0)/(1-\beta_0))$ is the transverse rapidity; 
$I_0$ and $I_1$ are modified Bessel functions. 
Such a distribution yields:
\begin{equation}
v_1=\frac{\beta_a p_t \cosh(y_t)}{2 T} 
   \left[ 1- \frac { I_1(\xi) E^2 \sinh(y_t) \cosh(y_t)  
     +I_0(\xi)(p_t T \cosh(y_t) - E T \sinh(y_t)^2)} 
               {p_t T \cosh(y_t) (\chi I_0(\xi) - \xi I_1(\xi))} \right].
\label{v1rel}
\end{equation}
Note, that the energy $E$ is measured in the frame moving longitudinally 
with the same velocity as an effective source. 
Studying particle production close to the fragmentation region, one should
remember, that the rapidity of the effective source could be less than that
of the particles~\cite{lmisk}. 
In this case $E=m_t \cosh(y^*-y)$, where
$y^*$ is the effective source rapidity. 
 
If the particles are emitted from a thermalized {\em spherical shell} at
temperature $T$, expanding with velocity $\beta_r$ (we use different
notation to distinguish $\beta_r$ from $\beta_0$, the expansion 
velocity in the transverse {\em plane}),
the expected invariant distribution has the form~\cite{lre,leos}:
\begin{equation}
J \propto \exp(-E \gamma /T) \left[\frac{\sinh(\alpha)}{\alpha}
(E +T/\gamma)-\cosh(\alpha) T/\gamma \right],
\end{equation}
where $\gamma=1/\sqrt{1-\beta_r^2}$, and $\alpha=\beta_r p \gamma /T$.
An application of the prescription~(\ref{ev1}) gives:
\begin{eqnarray}
v_1&=& \frac{p_t \beta_a \gamma}{2T}
    \left\{
      E \sinh(\alpha)/\alpha -\cosh(\alpha)T/\gamma+
    [(E+T/\gamma)(\cosh(\alpha)/\alpha 
    -\sinh(\alpha)/\alpha^2) 
    \right. 
\nonumber
\\& & \left.
    -\sinh(\alpha) T / \gamma]\beta_r E /p
    \right\} / \left\{
      (E +T/\gamma)\sinh(\alpha)/\alpha
    -\cosh(\alpha) T/\gamma \right\}
\end{eqnarray}
Qualitatively this model predicts the same dependencies $v_1(p_t)$ as
the model with the expansion only in the transverse plane. 
But the value of the expansion velocity needed in this model
for $v_1(p_t)$ exhibits a dip at low $p_t$ is about 15\% larger due to 
the 3-dimensional type of the expansion.

Summarizing, it is shown that a study of the dependence of the amplitude
of the first harmonic in the azimuthal distributions on transverse
momentum reveals important features of directed and transverse radial flow.
The magnitude and the shape of $v_1(p_t)$ is very sensitive to such 
parameters of the system as an effective temperature, radial expansion 
and directed flow velocities. 
Strong radial expansion (approximately $\beta_0 \geq \sqrt{2 T/ m}$) 
would evidence itself in the directed flow signal showing an anti-flow of 
the particles with low transverse momentum.
The preliminary results of E877 Collaboration presented at HIP-AGS~'96 
Workshop~\cite{lchang} suggests that such a regime could be the case in 
gold-gold collisions at the AGS.

The author acknowledge the discussions with W.-C. Chang.

\clearpage
\newpage

\section{Figure Caption}

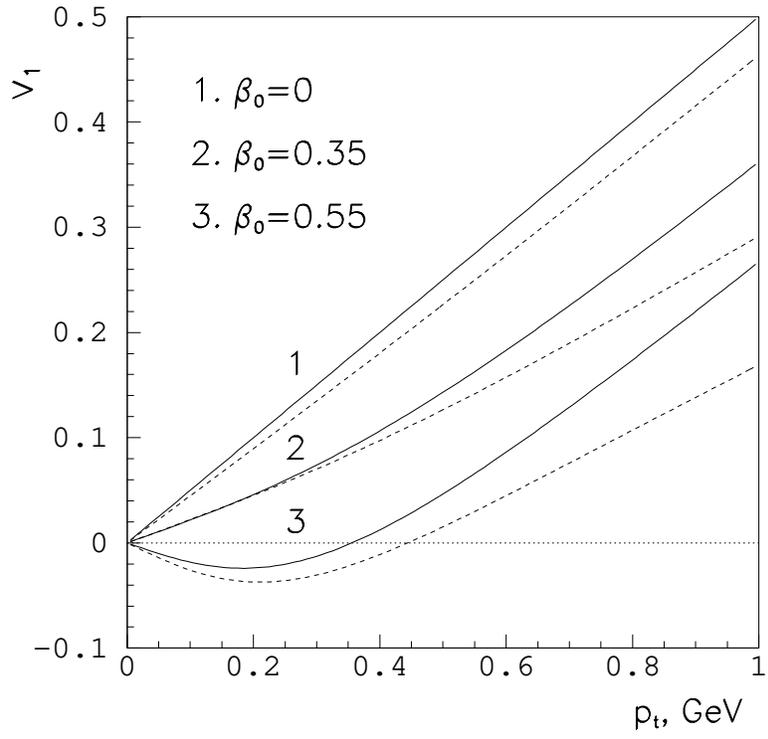
FIG.~1. $v_1(p_t)$: solid lines -- Eq.~(\ref{v1nr}), 
dashed lines -- Eq.~(\ref{v1rel}).
$T=0.1~GeV$, $\beta_a=0.1$.

\end{document}